\documentclass[twocolumn,amsmath,amssymb,nofootinbib,tighten,floatfix,prl]{revtex4}

\usepackage{color}
\usepackage{graphicx}
\usepackage{dcolumn} 
\usepackage{bm, bbm}      
\usepackage{curves}
\usepackage{epic}
\usepackage{wasysym}
\usepackage{epsf, times}
\usepackage{subfigure}
\usepackage{eufrak}
\usepackage{amsthm}

\begin{document}

\title{ 
Irrational {\it vs.} rational charge and statistics in
two-dimensional quantum systems }

\author{Claudio Chamon$^{1,2}$}
\author{Chang-Yu Hou$^{1}$}
\author{Roman Jackiw$^{2}$}
\author{Christopher Mudry$^{3}$}
\author{So-Young Pi$^{1}$}
\author{Andreas P.\ Schnyder$^{4}$}

\affiliation{$^{1}$\!\!\!
Physics Department, Boston University, Boston, MA 02215, USA
\\
$^{2}$\!\!\!
Department of Physics, Massachusetts Institute of Technology,
Cambridge, MA 02139, USA
\\
$^{3}$\!\!\!
Condensed matter theory group,
Paul Scherrer Institut,
CH-5232 Villigen PSI , Switzerland
\\
$^{4}$\!\!\!
Department of Physics, University of California, Santa Barbara, CA 93106, USA
}

\date{\today}

\begin{abstract}
We show that quasiparticle excitations with irrational charge
and irrational exchange statistics exist in tight-biding systems
described, in the continuum approximation, by the Dirac equation in
(2+1)-dimensional space and time. These excitations can be deconfined
at zero temperature, but when they are, 
the charge re-rationalizes to the value $1/2$ and
the exchange statistics to that of ``quartons'' (half-semions).

\end{abstract}

\maketitle

\def\openone{\leavevmode\hbox{\small1\kern-4.2pt\normalsize1}}

\newcommand{\slapar}{\not \hskip -2 true pt \partial\hskip 2 true pt}
\newcommand{\slapartxt}{\not\!\!\!\!\partial}
\newcommand{\slaA}{\!\not\!\! A}
\newcommand{\slaAA}{\!\not\!\! A^{5}}
\newcommand{\slaM}{\ \backslash \hskip -8 true pt M}
\newcommand{\slaS}{\ \backslash \hskip -7 true pt \Sigma}
\newcommand{\beq}{\begin{equation}}
\newcommand{\eeq}{\end{equation}}
\newcommand{\bea}{\begin{eqnarray}}
\newcommand{\eea}{\end{eqnarray}}
%
%

\textit{Introduction} -- It was shown by Jackiw and
Rebbi~\cite{Jackiw76} and by Su, Schrieffer, and Heeger~\cite{Su79}
that excitations with fermion number $1/2$ (or charge $1/2$) 
exist at domain walls in the dimerization pattern of electrons hopping
along a chain, as is believed to occur in polyacetylene.  For
electrons hopping on the honeycomb lattice, the lattice relevant to
graphene, a topological defect in a dimerization pattern that is
realized by a vortex was shown to lead to a topological zero-mode and
bind the fermion number $1/2$ to the vortex~\cite{Hou07}. Fermion-number
fractionalization in both polyacetylene and graphene can be understood
in terms of the spectral properties of one-dimensional (1D) and
two-dimensional (2D) massive Dirac Hamiltonians, respectively, that
describe the low energy limit of the electronic tight-binding
Hamiltonians. These fractionally charged topological excitations are
generically deconfined in 1D.  Their deconfinement in 2D relies on a
mechanism for the screening of the 2D Coulomb
potential by thermal~\cite{Hou07} or quantum fluctuations
involving an axial gauge field~\cite{Jackiw07}.

Applying different potentials to odd and even sites of the linear
chain results in a continuously varying fractional fermion number
\cite{Goldstone81,Rice82,Jackiw83}. At the level of the Dirac
Hamiltonian, this perturbation is represented by a second
(gap-opening) mass term, that adds in quadrature to the mass due to
the hopping dimerization of the chain. A complex order parameter is
constructed from these two masses as its real and imaginary pieces,
and the fractional fermion number is related to the phase twist of
this order parameter as it sweeps through a domain wall, a result
that has a natural interpretation within a bosonization scheme
\cite{Goldstone81}. Although exchange statistics is ill-defined in 1D,
this varying phase twist also implies a continuously varying exclusion
statistics \cite{Haldane91}.

Can the charge and the exchange statistics of fractionalized
quasiparticles in 2D be continuously varied as well? Here we show that
they can. The fermion numbers of quasiparticles bound
to a vortex can thus be \textit{irrational}. Remarkably, if an axial gauge
field supporting a half vortex is added to (precisely) screen the
interaction potential between quasiparticles, their fractional fermion
number {\it re-rationalizes} to the value $Q=1/2$ and their
statistical angle (for when time-reversal symmetry is broken) to the
value $\theta/\pi=1/4$.  These results are first derived at the level
of an effective field theory in $(2+1)$-dimensional space and time. We
then discuss the relevance of this analysis for planar tight-binding
models.

\textit{Definitions} -- 
The massive Dirac Lagrangian that we shall consider in this letter
takes the form
\begin{subequations}
\label{eq:def-massive-Dirac-Lagrangian}
\begin{equation}
\mathcal{L}:=
\bar{\Psi}
\left[
\gamma^{\mu}
\left(
{i}
\partial_{\mu}
+
\gamma^{\ }_{5}
A^{\ }_{5\mu}
\right)
-
M^{\ }_{a} \Delta^{\ }_{a}
\right]
\Psi,
\end{equation}
with 
$\mu=0,1,2$ and $a=0,1,2,3$, 
$\bar\Psi\equiv\Psi^\dagger\gamma^{0}$,
and $4\times 4$ matrices
\begin{equation}
\begin{split}
&
\gamma^{0}\equiv 
\left(\begin{array}{cc} 0 & {I} \\ {I} & 0 
\end{array}\right), 
\quad 
\gamma^{i} \equiv 
\left(\begin{array}{cc} 
0 & -\sigma_i \\ \sigma_i & 0
\end{array}\right),
\quad
\gamma^{\ }_{5}\equiv
{i}\gamma^{0}\gamma^{1}\gamma^{2}\gamma^{3},
\\
&
M^{\ }_{1}=\openone,
\quad
M^{\ }_{2}=-{i}\gamma^{\ }_{5}, 
\quad
M^{\ }_{3}=\gamma^{3},
\quad
M^{\ }_{0}=\gamma^{\ }_{5}\gamma^{3}.
\end{split}
\end{equation}
\end{subequations}
We allow for the background fields
$A^{\ }_{5\mu}$ and
$\Delta^{\ }_{a}$ to vary in space-time
$x\equiv(x^{\mu})\equiv(t,\boldsymbol{r})$.
The field
$A^{\ }_{5\mu}$ couples to the Dirac fermions
as an axial $U(1)$ gauge field does~\cite{Jackiw07}.
The four fields $\Delta^{\ }_{a}$, when constant in space and time, 
open an energy gap in the Dirac spectrum.
These masses have the following physical meaning on the 
honeycomb lattice, e.g., graphene. 
The masses ${\Delta^{\ }_{1,2}}$ correspond to
the two components of the complex Kekul\'e bond-density-wave order
parameter $\Delta\equiv\Delta^{\ }_{1}+{i}\Delta^{\
}_{2}$~\cite{Hou07}. The mass ${\Delta^{\ }_{3}}\equiv {\mu^{\
}_{\mathrm{s}}}$ is a staggered chemical potential that favors charges
to sit in one of the sublattices of the honeycomb
lattice~\cite{Semenoff84}. Finally, ${\Delta^{\ }_{0}}=\eta$ is the
only of the four masses that breaks time-reversal symmetry
(TRS). It originates on the lattice from a next-neighbor hopping term
with phases that was introduced in Ref.~\cite{Haldane88}. 
We will use a constant $\eta$ to discuss quasiparticle statistics.
The TRS masses $\mu^{\ }_{\mathrm{s}}$, $\Delta^{\ }_{1}$, and $\Delta^{\
}_{2}$ add in quadrature. 
Hence, we define the fields $m$ and 
$\boldsymbol{n}=(n^{\ }_{1},n^{\ }_{2},n^{\ }_{3})$ 
by
\begin{equation}
m:=\sqrt{\Delta^{2}_{1}+\Delta^{2}_{2}+\mu^{2}_{\mathrm{s}}},
\qquad
n^{\ }_{i}:=
\frac{\Delta^{\ }_{i}}{m}.
\label{eq: def n's}
\end{equation}

\textit{Induced fermionic $U(1)$-charge current with TRS} -- The current
$j^{\mu}(x)\equiv \langle \bar\Psi(x)\gamma^\mu \Psi(x)\rangle$ can be
computed perturbatively in a derivative expansion in 
$\boldsymbol{n}$.
When $A^{\mu}_{5}=0$, it reads
\begin{equation}
j^{\mu}=
\frac{1}{8\pi}\,
\epsilon^{\mu\nu\rho}\,
\boldsymbol{n}
\cdot
\left(
\partial^{\ }_{\nu}\boldsymbol{n}
\wedge
\partial^{\ }_{\rho}
\boldsymbol{n}
\right).
\label{eq:2d-topological-current}
\end{equation}
Equation~(\ref{eq:2d-topological-current}) is the desired extension to
2D of the bosonized current in a Luttinger liquid.  For a static
charge-1 vortex that vanishes at the origin and has the large distance
asymptotic form
\begin{equation}
\begin{split}
&
\Delta(\boldsymbol{r})=
\Delta(\infty)\,e^{{i}\phi}
+
\mathcal{O}(r^{-2}),
\\
&
\partial^{\ }_{i}\Delta(\boldsymbol{r})=
-
{i}
\Delta(\infty)
\frac{\epsilon^{\ }_{ij}r^{j}}{r}
e^{{i}\phi}
+
\mathcal{O}(r^{-3}),
\end{split}
\label{eq: def charge 1 vortex}
\end{equation}
where $\boldsymbol{r}=r(\cos\phi,\sin\phi)$,
we find the charge
\begin{equation}
Q=
\frac{1}{2}
\left(
1
-
\frac{
\mu^{\ }_{\mathrm{s}}(\infty)/\Delta(\infty)
     }
     {
\sqrt{
1
+
\mu^{2 }_{\mathrm{s}}(\infty)/\Delta^{2}(\infty)
     }
     }
\right)\;\;({\rm mod}\;1)
\label{eq:main-result-for-charge}
\end{equation}
that varies continuously with the asymptotic value
of the dimensionless ratio
$\mu^{\ }_{\mathrm{s}}(\infty)/\Delta(\infty)$.
The ambiguity mod 1 in the fermion number $Q$ 
corresponds to whether the bound state in the gap is empty or occupied. 
For $\mu^{\ }_{\mathrm{s}}(\infty)=0$,
$Q=\pm1/2$ is recovered as in Ref.~\cite{Hou07}.

When $A^{\mu}_{5}\neq0$,
the induced fermionic current is
$J^{\mu}=j^{\mu}_{\mathrm{cov}}+\delta j^{\mu}$
where $j^{\mu}_{\mathrm{cov}}$ is Eq.~(\ref{eq:2d-topological-current})
in which the derivatives have been replaced by 
the covariant derivatives
$
D^{\mu}=
\partial^{\mu}+2iA^{\mu}_{5}
$,
$
\delta j^{\mu}=
-
\frac{1}{2\pi}
\frac{\mu^{\ }_{\mathrm{s}}}{m}
F^{\mu}_{5},
$
and
$
F^{\mu}_{5}=e^{\mu\nu\rho}\partial^{\ }_{\nu}A^{\ }_{5\rho}.
$
The current $J^{\mu}$ is conserved.
The additional charge contributed by
$\delta j^{\mu}$
is proportional to the flux carried by $A^{\ }_{5\mu}$.
In fact, in the presence of the static vortex%
~(\ref{eq: def charge 1 vortex})
and of the static half-vortex
$
A^{i}_{5}(\boldsymbol{r})=
-\epsilon^{ij}\frac{r^{j}}{r^2}a^{\ }_{5}(r)
$
where $a^{\ }_{5}(0)=0$ and 
$a^{\ }_{5}(\infty)=1/2$,
the fractional charge is $Q=1/2$ 
for any $\mu^{\ }_{\mathrm{s}}(\infty)/\Delta(\infty)$!

Observe that, when $A^{\ }_{5\mu}=0$, despite the fact that the Dirac
Hamiltonian associated to Eq.~(\ref{eq:def-massive-Dirac-Lagrangian})
\textit{does not} possess an $SU(2)$ symmetry (the mass-matrices $M^{\
}_{1,2,3}$ do not close an $SU(2)$ algebra under multiplication), the
induced fermionic current~(\ref{eq:2d-topological-current}) is the
same as that in the $O(3)$ nonlinear sigma model (NLSM) derived from
Dirac Hamiltonians with an internal $SU(2)$
symmetry~\cite{Jaro84,Chen89}.  This is so for the following reason.
The nonunitary transformation $\bar\Psi=\bar\chi M^{\ }_{0}$ and
$\Psi= \chi$ is harmless as it induces a constant Jacobian that
cancels for any physical observable. Under this transformation, the
matrices $\Sigma^{\ }_{i}\equiv M^{\ }_{0}M^{\ }_{i}$ close the
$SU(2)$ algebra $[\Sigma^{\ }_{i},\Sigma^{\ }_{j}]= 2{i}\,\epsilon^{\
}_{ijk} \Sigma^{\ }_{k}$ and commute with the transformed matrices
$\Gamma^{\mu}=M^{\ }_{0}\gamma^{\mu}$.  The transformed Lagrangian%
~(\ref{eq:def-massive-Dirac-Lagrangian}) is then invariant under the
$SU(2)$ transformations generated by $\boldsymbol{\Sigma}= (\Sigma^{\
}_{1},\Sigma^{\ }_{2},\Sigma^{\ }_{3})$ when $A^{\mu}_{5}=0$.

\begin{figure}
\includegraphics[angle=0,scale=0.4]{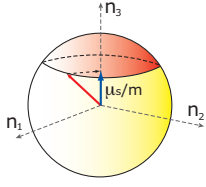}
\caption{
(color online). 
The 2 sphere spanned by the masses
$\boldsymbol{n}\equiv
(n^{\ }_{1},n^{\ }_{2},n^{\ }_{3})
=
m^{-1}(\Delta^{\ }_{1},\Delta^{\ }_{2},\mu^{\ }_{\mathrm{s}})$.
A vortex of charge 1 in the dimerizations
$\Delta^{\ }_{1}$ and $\Delta^{\ }_{2}$ corresponds to a parallel
on the sphere whose height is fixed by the 
asymptotic value of the staggered chemical potential 
$\mu^{\ }_{\mathrm{s}}(\infty)$.
The spherical area red (or yellow) enclosed by this parallel gives 
the fractional charge
induced by the fermionic zero mode localized at the core of the vortex.
        }
\label{fig:the-2-sphere}
\end{figure}

\textit{Exchange statistics} -- 
If two identical quasiparticles
associated with two vortices are exchanged, the
many-particle wavefunction changes by the phase $\exp({i}\theta)$
with $\theta$ the statistical angle.
On general grounds, we expect that fractionally charged quasiparticles
acquire a mutual fractional statistics. However, for quasiparticles to 
display fractional statistics one needs to break TRS, and thus we now include
a uniform
TRS-breaking 
$\Delta^{\ }_{0}\equiv\eta$. The calculation of the fractional
charge when $\eta=0$ has shown that fractionalization is of 
topological origin in that it only depends on
the asymptotic values of the fields. The same 
fractional charge follows when we impose the 
nonlinear constraint  
$
\boldsymbol{n}^{2}=n^{2}_{1}+n^{2}_{2}+n^{2}_{3}=1,
$
with $m$ in Eq.~(\ref{eq: def n's}) held constant
(see Fig.~\ref{fig:the-2-sphere}).
Now, a uniform $\eta$ competes with a uniform $m$ 
leading to a quantum critical point when $m=\eta$. 
Correspondingly, the fractional charge can be computed
for a nonvanishing ratio $\eta/m$
and shown to be the same as before
except for the multiplicative factor
$
\kappa^{\ }_{\mathrm{C}}:= \Theta(m-|\eta|)
$ 
with $\Theta$ the Heaviside step function. 
Moreover, for $A^{\mu}_{5}=0$, 
the effective action governing the dynamics
of the background fields $\boldsymbol{n}$ 
is, despite the lack of $SU(2)$ symmetry in
Eq.~(\ref{eq:def-massive-Dirac-Lagrangian}), 
the same as that of the
$O(3)$ NLSM of Refs.~\cite{Jaro84,Chen89}:
\begin{subequations}
\label{eq:NLSM}
\begin{equation}
S^{\ }_{\mathrm{eff}}=
\frac{m}{8\pi}
\int d^{3} x\,
\left(\partial^{\ }_{\mu}\boldsymbol{n}\right)^{2}
+
{i}\pi\kappa^{\ }_{\mathrm{H}}\,\mathrm{sgn}(\eta)
\;
S^{\ }_{\mathrm{Hopf}}
+
\dots
\end{equation}
The Hopf term is nonlocal in terms of the fields $\boldsymbol{n}$.
One possible representation is \cite{Hlousek90}
(see also Ref.~\onlinecite{Abanov00})
\begin{equation}
S^{\ }_{\mathrm{Hopf}}=
\frac{\epsilon^{\ }_{\mu\nu\rho}}{48\pi^{2}}
\int\! d^{3} x\,
\mathrm{tr}
\left[
\left(
U{i}\partial^{\mu}U^{\dag}
\right)
\left(
U{i}\partial^{\nu}U^{\dag}
\right)
\left(
U{i}\partial^{\rho}U^{\dag}
\right)
\right]
\label{eq:Hopf-term}
\end{equation}
\end{subequations}
with $U$ the unitary $4\times4$ matrix-field that rotates the
space-time dependent vector $\boldsymbol{n}$ into a fixed direction,
say $(0,0,1)$:
$\boldsymbol{\Sigma}\cdot\boldsymbol{n}=U^{\dag}\Sigma^{\ }_{3}U$.
When the angle $\kappa^{\ }_{\mathrm{H}}:=
\Theta(|\eta|-m)$ is nonvanishing, the angle 
$\kappa^{\ }_{\mathrm{C}}$ vanishes and vice versa.  
Both the fractional charge and the Hopf term
contribute to the statistical angle,
\begin{equation}
\begin{split}
\theta/\pi=&\,
\mathrm{sgn}(\eta)\;
Q^{2}
\Big(
\kappa^{\ }_{\mathrm{C}}
+
\kappa^{\ }_{\mathrm{H}}
\Big)
=\,
\,
\mathrm{sgn}(\eta)\;
Q^{2}
\end{split}.
\label{eq: final result theta}
\end{equation}

A heuristic derivation of~Eq.~(\ref{eq: final result theta})
when $m>|\eta|$ and $A^{\mu}_{5}=0$ is the following. 
Consider first a uniform $\boldsymbol{n}$, for which
the square of the Dirac Hamiltonian $\mathcal{H}$ associated to
Eq.~(\ref{eq:def-massive-Dirac-Lagrangian}) is
$
\mathcal{H}^{2 }=
\left(
\boldsymbol{p}^{2}
+
m^2
+
\eta^{2}
\right)
+
2\eta
\sum_{i=1}^{3}
\Sigma^{\ }_{i} n^{\ }_{i}.
$
This Hamiltonian resembles that of a particle with spin in a magnetic
field, of strength proportional to $\eta$, 
that points in the $\boldsymbol{n}$ direction (hereafter, we take $\eta>0$).
If the uniform $\boldsymbol{n}$ is
rotated adiabatically by $2\pi$ around the $\hat{z}$ direction, a Berry phase $\gamma^{\ }_{\rm rot}(\boldsymbol{n})=2\pi\sin^2\frac{\alpha}{2}$ is accumulated for each filled single-particle electronic state, where $\alpha$ is the angle
that $\boldsymbol{n}$ makes with the north pole of the unit sphere.
The many-body Berry phase is the sum over occupied single-particle
Berry phases after subtraction of the Berry phase of some reference
many-body state ({\rm i.e}., without vortex).  Now, $\boldsymbol{n}$
is not spatially uniform for a vortex, and thus if all
$\boldsymbol{n}(\boldsymbol{r})$ are rigidly rotated by $2\pi$, their
accumulated phase depends on the texture.  In a semiclassical
approximation, the Berry phase at each $\boldsymbol{r}$ must be
weighted by the fermion density: $ \gamma^{\ }_{\rm rot}= \int
d^2\boldsymbol{r}\,j^{0}(\boldsymbol{r})\,2\pi
\sin^2\frac{\alpha(\boldsymbol{r})}{2} $.  With the help of
Eq.~(\ref{eq:2d-topological-current}) when $m>\eta$, this gives
$
\gamma^{\ }_{\rm rot}= 
\int\frac
{d\Omega}{4\pi}\,2\pi\sin^2\frac{\alpha}{2}= 
\pi\sin^4\frac{\alpha}{2}=
\pi\,Q^2.
$
Hence, the phase accumulated by
spinning the vortex by $2\pi$ is $\pi\,Q^2$,
which is what an object with spin $Q^2/2$
collects.  The exchange statistics for two quasiparticles of charge
$Q$ each should then be $\theta/\pi =
Q^2$.

\begin{figure}[t]
\includegraphics[angle=0,scale=0.6]{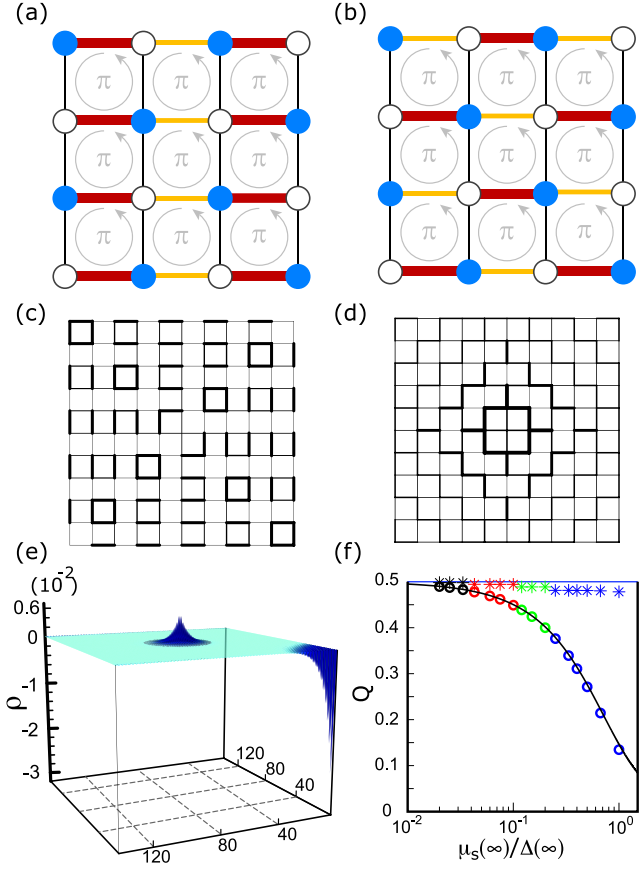}
\caption{
(color online). 
(a) The columnar dimerization pattern,
as indicated by the coloring of the bonds, 
for the nearest-neighbor hopping on a square lattice
in the background of the $\pi$-flux phase
that opens up the gap $\Delta(\infty)$.
A staggered chemical potential,
as indicated by the distinction between sites
of sublattice  $A$ and $B$
opens up the gap $\mu^{\ }_{\mathrm{s}}(\infty)$.
(b) The staggered dimerization pattern
for the nearest-neighbor hopping on a square lattice
that generates an axial gauge field.
(c) A charge-1 vortex in the columnar dimerization (a).
(d) A charge-1/2 vortex in the staggered dimerization (b)
with core radius $c=0.3$.
(e) The fermion density profile of (c) for a square lattice
with open boundary conditions and $144\times144$ sites.
(f) The fermion number as a function of the scaling variable
$\mu^{\ }_{\mathrm{s}}(\infty)/\Delta(\infty)$
in the presence of the single charge-1 vortex (c) 
or with the addition of the axial charge-1/2 vortex (d)
with core radius $c=0.01$. The staggered chemical potential $\mu^{\ }_{\mathrm{s}}$ takes the values $0.01t$ (black), $0.03 t$ (red), $0.06 t$ (green), and $0.1t $ (blue). The thick and thin lines are the prediction~(\ref{eq:main-result-for-charge}) without the axial vortex and 
$Q=1/2$, 
respectively.
        }
\label{fig:numerics}
\end{figure}

\textit{Numerics} -- We compare the predictions that the charge
varies continuously as a function of the scaling variable
$\mu^{\ }_{\mathrm{s}}(\infty)/\Delta(\infty)$ 
according to Eq.~(\ref{eq:main-result-for-charge}) 
in the absence of the axial vortex [$a^{\ }_{5}(\infty)=0$] 
and that the charge re-rationalizes to
$Q=1/2$ in the presence of the axial half-vortex 
[$a^{\ }_{5}(\infty)=1/2$]
to results from an exact diagonalization in a lattice model. The model
we consider is the $\pi$-flux phase \cite{Lieb94} for electrons
hopping on a square lattice, to which we add a dimerization pattern of
the hopping amplitudes that realizes, on the lattice, the mass and
axial vortices of Refs.~\cite{Hou07,Jackiw07}. We then study the
charge bound to these singularities as a function of the staggered
chemical potential $\mu^{\ }_{\mathrm{s}}$ for the cases 
$a^{\ }_{5}(\infty)=0,1/2$ (see also \cite{Seradjeh07}).

Consider a square lattice as in Fig.~\ref{fig:numerics} on which
spinless fermions hop. The square lattice can be divided into two
interpenetrating sublattices $A$ (open circles in
Fig.~\ref{fig:numerics}) and $B$ (filled circles in
Fig.~\ref{fig:numerics}), such that all nearest neighbors of sites in
$A$ belong to $B$, and vice versa. Here we construct the tight-binding
Hamiltonian as
$
H=
-
\sum_{r\in A}
\sum_{j=1}^4\; 
\left(t^{\pi}_{r,j}+\delta t^{\ }_{r,j}\right) 
a^{\dag}_{r}\,b^{\ }_{r+s^{\ }_{j}} 
+ 
\mathrm{H.c.},
$
where  $r=(m_1,m_2) \in \mathbbm{Z}^2$ labels the sites in $A$,
located at positions  $(m^{\ }_{1}+m^{\ }_{2})\hat{\mathrm x}
+(m^{\ }_{1} - m^{\ }_{2})\hat{\mathrm y}$
in the square lattice. The $s^{\ }_{j}$ are the four vectors (labeled
counterclockwise, starting from the $+\hat{\mathrm x}$ direction, by
$j=1,2,3,4$) connecting a site in $A$ to one of its four
nearest-neighbor sites in $B$.

The kinetic energy in Eq.~(\ref{eq:def-massive-Dirac-Lagrangian})
follows from linearizing at half-filling the tight-binding dispersion
when the only nonvanishing hopping amplitudes in units of $t>0$ are
$
t^{\pi}_{r,j}=
(-1)^{(m^{\ }_{1}+m^{\ }_{2})
(\delta^{\ }_{j,1}+\delta^{\ }_{j,3})}\,t.
$
These hoppings define the $\pi$-flux phase (they are gauge equivalent
to the case of uniform hoppings $t$ but with a magnetic flux of $\pi$
in units of $h/e$ threading each elementary plaquette, which is
indicated in Fig.~\ref{fig:numerics}).

A staggered chemical potential $+\mu^{\ }_{\mathrm{s}}$ on sublattice 
$A$ and $-\mu^{\ }_{s}$ 
on sublattice $B$ induces the bilinear $\bar{\Psi}M^{\ }_{3}\Psi$ in
Eq.~(\ref{eq:def-massive-Dirac-Lagrangian}) \cite{Semenoff84}. 
A dimerization, such as shown in Fig.~\ref{fig:numerics}(a), 
arises from  
$\delta t^{\ }_{r,j}\propto
(-1)^{m^{\ }_{1}+m^{\ }_{2}}
(\delta^{\ }_{j,1}-\delta_{j,3})t^{\pi}_{r,j}$.
Such dimerizations induce the mass bilinears 
$\bar{\Psi}M^{\ }_{1,2}\Psi$ 
in Eq.~(\ref{eq:def-massive-Dirac-Lagrangian}) ~\cite{Hou07}. 
A dimerization, such as shown in Fig.~\ref{fig:numerics}(b), 
arises from
$\delta t^{\ }_{r,j}\propto
(\delta^{\ }_{j,3}- \delta^{\ }_{j,1})\,t^{\pi}_{r,j}$.
Such dimerizations induce the axial bilinears 
$\bar{\Psi}\gamma^{\ }_{1,2}\gamma^{\ }_{5}\Psi$ 
in Eq.~(\ref{eq:def-massive-Dirac-Lagrangian}).  

A charge-$n$ vortex in the columnar dimerization pattern is defined by
\begin{equation}
\begin{split}
&
\delta t^{\ }_{r,j}=
t^{\pi}_{r,j}
\frac{\Delta(\infty)}{2t}
\left(
\cos n\theta
\cos\phi^{\ }_{j}
-
\sin n\theta
\sin\phi^{\ }_{j}
\right),
\\
&
\cos  \theta =
\frac{m^{\ }_{1}+m^{\ }_{2}}{|r|},
\qquad
|r|=\sqrt{2(m^{2}_{1}+m^{2}_{2})},       
\\
&
\phi^{\ }_{j}=
\pi(m^{\ }_{1}+m^{\ }_{2})
+
j(\pi/2),
\end{split}
\end{equation}
and shown in Fig.~\ref{fig:numerics}(c) when $n=1$.
A half vortex with core radius $c$
in the staggered dimerization
pattern is defined by
\begin{equation}
\delta t^{\ }_{r,j}= 
\frac{t^{\pi}_{r,j}}{2|r|}
\tanh\frac{|r|}{c}
\big[
(\delta^{\ }_{j,3}-\delta^{\ }_{j,1})
\cos\theta
+ 
(\delta^{\ }_{j,4}-\delta^{\ }_{j,2}) 
\sin\theta
\big]
\end{equation}
and shown in Fig.~\ref{fig:numerics}(d).   

The charge-density profile is obtained by adding the contributions from
all exact single-particle eigenstates that are filled: negative energy
states plus the bound state with energy within the gap that appears as
a consequence of the vortex. The fermionic density profile of the
vortex from Fig.~\ref{fig:numerics}(c) is shown in
Fig.~\ref{fig:numerics}(e) for a lattice made of $144\times144$ sites
with a weight concentrated either at the core of the vortex or on the
boundary (open boundary conditions are used).  The fermion number in
the presence of a vortex is approximated by integrating the local
fermion density in a disk surrounding the vortex that extends beyond
the localization length of the induced bound state but remains
insensitive to the fermion density that has accumulated at the
boundary. Subtraction of the background fermion charge in the absence
of the vortex is always implicit.  The continuous dependence of the
fermion number~(\ref{eq:main-result-for-charge}) on the scaling
variable $\mu^{\ }_{\mathrm{s}}(\infty)/\Delta(\infty)$ is shown in
Fig.~\ref{fig:numerics}(f) for the charge 1 vortex of
Fig.~\ref{fig:numerics}(c).  Also shown in Fig.~\ref{fig:numerics}(f)
is the effect of the axial charge-1/2 vortex from
Fig.~\ref{fig:numerics}(d) superimposed to the charge 1 vortex of
Fig.~\ref{fig:numerics}(c); as anticipated
we find that the fermion number stays close to $1/2$.
We conclude that the agreement between the field theory and the
numerics is good and improves as both 
$\mu^{\ }_{\mathrm{s}}(\infty)/\Delta(\infty)\to0$ 
and $\Delta(\infty)/t\to0$.

\textit{Energetics} -- At the level of the field theory,
the energy cost to separate vortices of opposite charge grows like the
logarithm of their separation. Temperature fluctuations screen
this interaction through the Kosterlitz-Thouless mechanism.
Alternatively, it was shown in Ref.~\cite{Jackiw07} that the bare
interaction between vortices is screened at zero temperature by
coupling them to an axial gauge field $A^{\ }_{5\mu}$
carrying a half-vortex.
The issue of energetics on the lattice is more subtle than in the
continuum approximation. The axial $U(1)$ symmetry of the field theory%
~(\ref{eq:def-massive-Dirac-Lagrangian}) is reduced to the finite
group $\mathbb{Z}^{\ }_{4}$ of rotations by $\pi/2$ about a site of
the square lattice when the dimerization pattern responsible for the
gap at the Fermi energy is commensurate with the lattice.  If so, the
energy cost to create two vortices of opposite charge grows
linearly with their separation. However, slight deformations of the
hopping amplitudes away from the $\mathbb{Z}^{\ }_{4}$-symmetric ones
move the position of the Dirac points in the Brillouin zone, rendering
the wavevector for the gap-opening dimerization pattern (which
connects the two Dirac points) incommensurate with the lattice while
preserving the bound state in the gap \cite{Hou07}. This
incommensuration can restore an energy cost for the creation of a
vortex-antivortex pair with a logarithmic dependence on their
separation, as is believed to happen in quantum dimer models, in the
mechanism termed Cantor deconfinment in Ref.~\cite{Fradkin04}.

\textit{Summary} -- We have calculated the irrational fermion number
attached to vortices in a complex-valued Higgs field coupled to
massive Dirac fermions in $(2+1)$-dimensional space time. We also
calculated the exchange statistics of these vortices and showed that
it is irrational as soon as the fractional fermion number
is. Deconfinement of the vortices at zero temperature requires the
(axial gauge) coupling to a half vortex~\cite{Jackiw07}. Remarkably,
this coupling re-rationalizes the fermion number to the value $1/2$.
We have compared our predictions for the induced fermion number with
an exact diagonalization study of a planar tight-binding model with a
static defective dimerization of the hopping amplitude and found a
good agreement.

This work is supported in part by Grants NSF DMR-0305482 (C.~C. and
C-Y.~H.), DOE DE-FG02-05ER41360 (R.~J.) and DOE DE-FG02-91ER40676
(S-Y.~P.).  A.~S. thanks the Swiss NSF for its financial support.  Use
of the CNSI Computer Facilities at UCSB is gratefully acknowledged.

\end{document}